\documentclass{cimento}

\title{Milagro Search for VHE Emission from GRBs in the Swift Era}
\author{P.~M.~Saz Parkinson\from{ins:x} for the Milagro Collaboration}
\instlist{\inst{ins:x} Santa Cruz Institue for Particle Physics,
       U.C.S.C., 1156 High Street, Santa Cruz, CA 95064}
\PACSes{\PACSit{98.70.Rz}{ gamma-ray sources; gamma-ray bursts}}
\begin{document}

\maketitle

\begin{abstract}
Since its launch, in late 2004, Swift has been locating gamma-ray
bursts (GRBs) at a rate of $\sim$100 per year. 
Very high energy (VHE) emission ($>$100 GeV) is predicted by
several models. Here, we present the results of a search for 
VHE emission from the most recent GRBs to fall within the Milagro
 field of view.

\end{abstract}


Milagro is a wide field (2 sr), high duty cycle ($>$ 90\%),
ground-based water Cherenkov gamma-ray telescope which monitors the northern 
sky almost continuously from 0.1--100 TeV~\cite{atkins00}. At 
these energies, gamma rays are attenuated by the redshift-dependent extra-galactic 
background light (EBL)~\cite{primack05}, making GRBs above z$>$0.5 very hard to detect. Milagro
has been operating (and searching for VHE emission from GRBs) since 2000~\cite{2005ApJ...630..996A}.



The launch of {\em Swift} has increased 
greatly the number of well-localized GRBs. Here 
we present the results of a search for an excess of events above those due to the background 
for 39 GRBs detected by several satellites (primarily {\em Swift}) between December 2004 and May 
2006. Table~\ref{grb_table} lists the GRBs in the sample and summarizes the results. The number 
of events falling within a 1.6 degree bin is summed for the relevant duration 
(column 2 of Table~\ref{grb_table}) and the number of background events is estimated from two 
hours of data surrounding the burst, using a technique known as ``direct integration''~\cite{atkins03b}. 
No significant emission was detected from any of the locations. We present 
upper limits on the fluence in column 7 of Table~\ref{grb_table}. For those bursts with known redshift, 
we compute the effect of EBL absorption, according to the model of ref.~\cite{primack05} and 
print the upper limits in bold. 


\begin{table}
\begin{tabular}{llllllll}
\hline
GRB & T90/Dur. & Zenith angle, $\theta$ & z & Instrument & $\sigma$ & 99\% UL(fluence) \\
\hline
041219a & 520   & 26.9  & ... 	& INTEGRAL 	 	& +1.7 & 5.8e-6  \\
050124  & 4    	& 23.0  & ... 	& Swift		  	& -0.8 & 3.0e-7  \\
050319  & 15   	& 45.1  & 3.24 	& Swift		  	& +0.6 & ...  \\
050402  & 8   	& 40.4  & ... 	& Swift		  	& +0.6 & 2.1e-6  \\
050412  & 26   	& 37.2  & ... 	& Swift	 	  	& -0.6 & 1.7e-6  \\
050502  & 20   	& 42.7  & 3.793 & INTEGRAL	 	& +0.6 & ...  \\
050504  & 80   	& 27.6  & ... 	& INTEGRAL	  	& -0.8 & 1.3e-6  \\
050505  & 60   	& 28.9  & 4.3 	& Swift		  	& +1.2 & ...  \\
050509b & 0.128 & 10.0  & 0.226?& Swift	 	  	& -0.9 & \textbf{1.1e-6} \\
050522  & 15   	& 22.9  & ... 	& INTEGRAL	  	& -0.6 & 5.1e-7  \\
050607  & 26.5  & 29.3  & ... 	& Swift	 	 	& -0.9 & 8.9e-7  \\
050712  & 35   	& 38.8  & ... 	& Swift	 	  	& -0.1 & 2.5e-6  \\
050713b & 30   	& 44.2  & ... 	& Swift	 	  	& -0.3 & 4.0e-6  \\
050715  & 52   	& 36.9  & ... 	& Swift	 	  	& -1.5 & 1.7e-6  \\
050716  & 69   	& 30.3  & ... 	& Swift	 	  	& -0.5 & 1.6e-6  \\
050820  & 20   	& 21.9  & 2.612 & Swift	 	 	& +0.2 & ...  \\
051103  & 0.17  & 49.9 	& 0.001?& IPN		& -0.2 & \textbf{4.2e-6} \\ 
051109  & 36    & 9.7   & 2.346 & Swift		  	& -1.1  & \textbf{4.3e-3}  \\
051111  & 20   	& 43.7  & 1.55 	& Swift		  	& +0.7 & \textbf{3.8e-2}  \\
051211b & 80	& 33.3	& ...	& INTEGRAL		& +0.4	& 2.6e-6	\\
051221	& 1.4	& 41.8	& 0.55	& Swift			& +0.6	& \textbf{9.8e-4} \\
051221b	& 61	& 25.9	& ... 	& Swift			& +1.5	& 1.8e-6 \\
060102	& 20	& 39.9	& ... 	& Swift			& -0.9	& 2.0e-6 \\
060109	& 10	& 22.4	& ... 	& Swift			& -1.3	& 4.1e-7 \\
060110	& 15	& 43.0	& ... 	& Swift			& -0.3	& 3.0e-6 \\
060111b	& 59	& 36.5	& ... 	& Swift			& -0.6	& 2.3e-6 \\
060114	& 100	& 40.6	& ... 	& INTEGRAL		& +0.5	& 5.1e-6 \\
060204b	& 134	& 30.5	& ... 	& Swift			& +0.3	& 2.7e-6 \\
060210	& 5	& 43.4	& 3.91 	& Swift			& +0.6	& 2.9e-6 \\
060218$^{*}$	& 10	& 44.6	& 0.03 	& Swift			& +2.4	& \textbf{3.8e-5} \\
060306	& 30	& 46.2	& ... 	& Swift			& +1.0	& 7.2e-6 \\
060312	& 30	& 43.6	& ... 	& Swift			& -1.0	& 3.3e-6 \\
060313	& 0.7	& 46.7	& ... 	& Swift			& -0.5	& 2.7e-6 \\
060403	& 25	& 27.6	& ... 	& Swift			& -0.1	& 1.0e-6 \\
060427b	& 0.2	& 16.4	& ... 	& IPN			& +0.6	& 1.7e-7 \\
060428b	& 58	& 26.6	& ... 	& Swift			& -1.1 	& 1.1e-6 \\
060507	& 185	& 47.1	& ... 	& Swift			& +0.4	& 1.6e-5 \\
060510b	& 300	& 42.8	& 4.9 	& Swift			& +1.9	& ... \\
060515	& 52	& 41.5	& ... 	& Swift			& -0.3	& 9.6e-6 \\

\hline
\end{tabular}
\caption{List of GRB in the field of view of Milagro in the Swift Era (December 2004 -- May 2006), with 
preliminary 99\% confidence upper limits on the fluence (0.2--20 TeV), in ergs cm$^{-2}$.
$^{*}$This burst, due to its long duration of more than 2000 s left Milagro's field of view while in 
progress. The limit presented here is for the 10 s hard spike reported by the instrument team.}\label{grb_table}
\end{table}


\acknowledgments
We have used GCN Notices to select raw data for archiving and use in this search, and
we are grateful for the hard work of the GCN team, especially Scott Barthelmy. We acknowledge 
Scott Delay and Michael Schneider for their dedicated efforts in the construction and 
maintenance of the Milagro experiment.This work has been supported by the National Science 
Foundation, 
the US Department of Energy (Office of High-Energy Physics and 
Office of Nuclear Physics), Los Alamos National Laboratory, the University of
California, and the Institute of Geophysics and Planetary Physics.  
I appreciate the support of the American Astronomical Society and the 
National Science Foundation in the form of an International Travel Grant, 
which enabled me to attend this conference. I am also grateful to the
conference organizers for their financial support.

\def \atel {The Astronomer's Telegram}
\def \apj {ApJ}
\def \aj {AJ}
\def \apjl {ApJL}
\def \mnras {MNRAS}
\def \iaucirc {IAUCIRC}
\def \em { }
\def \aap {A\&A}
\def \nat {Nature}
\def \araa {Anual Review of Astronomy and Astrophysics}

\end{document}